\documentclass[
a4paper,
aps,
prd,
twocolumn,
superscriptaddress,
preprintnumbers,
]{revtex4-1}

\usepackage{graphicx}
\usepackage{dcolumn}
\usepackage{bm}

\usepackage{psfrag}
\usepackage{color}

\begin{document}

\preprint{CPHT-024.0411}
\preprint{UAB-FT-692}

\title{Improving naturalness in warped models with a heavy bulk Higgs boson}

\author{Joan A. Cabrer}
%
\affiliation{%
Institut de F\'isica d'Altes Energies, Universitat Aut{\`o}noma de Barcelona, 08193 Bellaterra, Barcelona, Spain
}%


\author{Gero von Gersdorff}
\affiliation{
Centre de Physique Th\'eorique, \'Ecole Polytechnique and CNRS, F-91128 Palaiseau, France
}%

\author{Mariano Quir\'os}
\affiliation{
Instituci\'o Catalana de Recerca i Estudis  
Avan\c{c}ats (ICREA) and 
Institut de F\'isica d'Altes Energies, 
Universitat Aut{\`o}noma de Barcelona, 08193 Bellaterra, Barcelona, Spain
}%



\begin{abstract}
A Standard-Model-like Higgs boson should be light in order to comply with
electroweak precision measurements from LEP. We consider five-dimensional (5D) warped models -- with a deformation of the metric in the IR region -- 
as UV completions of the Standard Model with a heavy Higgs boson. Provided the Higgs boson propagates in the 5D bulk the Kaluza Klein (KK) modes of the gauge bosons can compensate for the Higgs boson contribution to oblique parameters while their masses lie within the range of the LHC. The little hierarchy between KK scale and Higgs mass essentially disappears and the naturalness of the model greatly improves with respect to the AdS (Randall-Sundrum) model. In fact the fine-tuning is better than 10\% for all values of the Higgs boson mass.
\end{abstract}

\maketitle


The Standard Model (SM) of electroweak interactions suffers from a naturalness problem as the Higgs boson mass is sensitive to ultraviolet (UV) physics at the scale $\Lambda$ through quadratically divergent radiative corrections. In particular, the coupling of the top quark to the Higgs boson generates a one-loop shift on the Higgs boson mass which behaves as $\Delta m_H^2\sim (3/4\pi^2)\Lambda^2$, which translates into a sensitivity to the cutoff $\delta=3\Lambda^2/(4\pi^2m_H^2)$. Naturalness of the theory, i.e.~$\delta \sim 1$, then requires that $\Lambda\sim 3.6\, m_H$, which in turn implies for a light Higgs boson that the UV physics should be around the corner at LHC (e.g.~for $m_H\sim 115$~GeV, $\Lambda\sim 400$~GeV) while for a heavy Higgs boson the UV physics can be at much higher scales (e.g.~for $m_H\sim 600$~GeV, $\Lambda\sim 2.2$~TeV). 
In view of the negative results on new physics searches at LEP2, and the increasing bounds imposed by ongoing LHC searches, it is thus interesting to consider possible solutions to the hierarchy problem with SM UV completions able to accommodate a heavy Higgs boson
. 

The idea of improving the naturalness of the SM with a heavy Higgs boson is not a new one~\cite{Peskin:2001rw,Barbieri:2006dq}. Of course, since the SM is consistent with all electroweak precision tests (EWPT) for a light Higgs boson, it is necessary to introduce new physics to compensate for the contribution of a heavy Higgs boson~\cite{foot:2}. In this Letter we will consider heavy Higgs boson models where EWPT are saved by the UV physics solving the hierarchy problem. Since a heavy Higgs boson contributes negatively to the $T$ parameter, an obvious requirement is the presence of new states that violate custodial symmetry in such a way as to give a positive contribution to $T$. Such states are naturally provided by the KK modes of the hypercharge gauge boson in five dimensional (5D) warped compactifications. Models where the 5D metric is AdS were originally proposed by Randall and Sundrum (RS)~\cite{Randall:1999ee} to solve the hierarchy problem while fulfilling all experimental bounds~\cite{Davoudiasl:2009cd}. Furthermore, we have recently proposed~\cite{Cabrer:2010si} a class of models which departs from AdS in the IR and where electroweak precision observables (EWPO) are naturally suppressed, giving rise to milder bounds on the KK mode masses than in RS models. In particular, we will consider models in which the Higgs boson is propagating in the bulk. Considering for RS a heavy Higgs boson localized in the IR boundary was already proposed in Ref.~\cite{Casagrande:2008hr}, leading to KK masses $m_{KK}\gtrsim $ 8 TeV for a 450 GeV Higgs boson. In this Letter we will first prove that delocalizing the Higgs boson in the bulk one can lower the previous bound to $m_{KK}\gtrsim $ 4.6 TeV. Moreover, by considering more general metrics with conformal deformations in the IR one can obtain spectra of KK modes with masses in the LHC range.   

We will then consider the SM propagating in a 5D space with an arbitrary metric $A(y)\equiv A(ky)$, where $k$ is a constant with dimension of mass of order the Planck scale (identified with the curvature radius for the AdS case), such that in proper coordinates $ds^2=e^{-2A(y)}\eta_{\mu\nu}dx^\mu dx^\nu+dy^2$, $\eta_{\mu\nu}=(-1,1,1,1)$; and two UV and IR boundaries located at the edges of a finite interval at $y=0$ and $y=y_1$ respectively. We will consider 5D gauge fields $W^i_M(x,y)$ and $B_M(x,y)$ ($M=\mu,5$) propagating in the bulk, a stabilizing field $\phi(x,y)$ fixing the value of $A(y_1)$, as well as a bulk SM Higgs field
\begin{equation}
H(x,y)=\frac{1}{\sqrt 2}e^{i \chi(x,y)} \left(\begin{array}{c}0\\h(y)+\xi(x,y)
\end{array}\right)
\ ,
\label{Higgs}
\end{equation}
where $\chi(x,y)$ contains the 4D Goldstone bosons while $h(y)$ is the 5D Higgs background and $\xi(x,y)$ describes the Higgs field fluctuations. For the moment we will consider an arbitrary metric $A(y)$ and Higgs background $h(y)$.

As it has been shown in Ref.~\cite{Cabrer:2010si}, the parameters of the effective SM-like Lagrangian for the Higgs boson,
\begin{equation}
\mathcal L_{\rm eff}=-|D_\nu   \mathcal H|^2+\mu^2 |\mathcal H|^2-\lambda |\mathcal H|^4\,,
\end{equation}
behave as
\begin{equation}
\mu^2\sim Z^{-1} \,\rho^2 \,,\qquad \lambda\sim Z^{-2} \,,
\end{equation}
where we have suppressed $\mathcal O(1)$ coefficients given by the bulk and IR--boundary potentials.
The IR scale is defined as $\rho= k e^{-A(y_1)}$ and the dimensionless quantity $Z$,
\begin{equation}
Z=
k\int_0^{y_1}dy\frac{h^2(y)}{h^2(y_1)}e^{-2A(y)+2A(y_1)}\,,
\label{Z}
\end{equation}
depends on both the gravitational and Higgs backgrounds.  Notice that the physical Higgs boson mass is $\nobreak{m_H^2=2\mu^2\sim 2Z^{-1}\rho^2}$. This shows that in models with $Z=\mathcal O(1)$, as it is the case of the RS model, a heavy Higgs boson mass is more natural than a light one: its preferred value is $\rho$ unless a (little) fine-tuning in the bulk and brane potentials  is done. As we will see later on, there are models which depart from AdS in the IR region, where the condition $Z\gg 1$ can be satisfied and one can easily accommodate at tree level a light Higgs boson. Nevertheless, in all cases radiative corrections in the effective theory below the scale $\Lambda\sim m_{KK}$ of order a few TeV will tend to destabilize light Higgs boson masses. Hence, some degree of fine-tuning is needed to keep the Higgs field light, while a heavy Higgs boson can more naturally be stable against radiative corrections. More details will be elaborated later on.  
 
We now make the KK-mode expansion $A_\mu(x,y)=a_\mu\cdot f_A(y)/\sqrt{y_1}$ where $A=A^\gamma,Z,W^\pm$ and the dot product denotes the expansion in modes, the functions $f_A$ satisfy the equations of motion
\begin{equation}
f_n''(y)-2A'(y)f_n'(y)+m_n^2e^{2A(y)}f_n(y)=0 \,,
\end{equation}
along with the Neumann boundary conditions $f_n'(y_i)$=0 and the mass eigenvalue $m_n$ is the mass of $a_\mu^n$. We adopt the normalization convention $\nobreak{\int_0^{y_1}f_n^2(y)dy=y_1}$.
The impact of the KK modes on the EWPO will depend crucially on how strongly the former couple to the Higgs currents. Writing the coupling as $\mathcal L = \sum_n \alpha_n(g\,W_\mu^n j_\mu^L+g'\,B_\mu^n j_\mu^Y) $,
we can express these couplings in terms of the $Z$ factors as
\begin{equation}
\alpha_n=\frac{k}{ Z}\int_0^{y_1}dy \frac{h^2(y)}{h^2(y_1)}e^{-2A(y)+2A(y_1)}f_n(y)\,.
\label{couplingKK}
\end{equation}
Enhanced $Z$ factors will reduce these couplings, provided the integrals stay approximately constant. A useful approximation for the KK wave function is given in terms of Bessel functions (up to normalization)
\begin{equation}
f_n(y)=z\left[Y_0(m_n z_0)J_1(m_nz)
-J_0(m_nz_0)Y_1(m_nz)
\right]Ê\,,
\end{equation}
where $z(y)-z(0)\equiv\int^y_0e^A\approx e^A/A'$ are the conformal coordinates. Note  that $A'$ can become large and hence $e^A$ can be very different from $z$.

Although the effective SM below a multi-TeV cutoff is more natural with a heavy Higgs boson, the present EWPT point towards a light Higgs boson. In particular, a $\chi^2$ fit of all SM EWPO drives the 95\% CL upper limit of the Higgs boson mass to $m_H\lesssim 150$ GeV. This means that a heavy Higgs boson needs to be accompanied by new physics to restore agreement with EWPT. In fact, from this point of view we can consider the Higgs boson mass measurement at LHC as a good test for new physics: if LHC found a heavy Higgs boson the SM would be excluded and new physics would be required, even if the Higgs field is not detected, motivating an  upgrade of the LHC and/or the construction of other colliders. In the rest of this Letter we will contrast warped models with heavy Higgs bosons and EWPT leading to lower bounds on KK mode masses, $m_{KK}$. For simplicity we will consider fermions localized in the UV brane such that the deviations in electroweak precision measurements are encoded in the momentum dependence of the propagators of the electroweak gauge bosons, or ``oblique corrections". 

Electroweak precision measurements are commonly mapped to the three Peskin-Takeuchi ($T,S,U$) parameters~\cite{Peskin:1991sw}, although in models with a gap between the electroweak and new physics scale $U$ is expected to be small and it has been suggested to instead consider the set $(T,S,W,Y)$
~\cite{Barbieri:2004qk}.
In the class of warped models we have just described, these observables are given by~\cite{Cabrer:2010si}
\begin{eqnarray}
\alpha T&=&s_W^2 m_Z^2 y_1\int_0^{y_1} e^{2A(y)}[1-\Omega(y)]^2
\,,\nonumber\\
\alpha S&=&8s_W^2c_W^2 m_Z^2 \int_0^{y_1} e^{2A}\left(y_1-y\right)[1-\Omega(y)]
 \,,\nonumber\\
Y&=&W=\frac{c_W^2m_Z^{2}}{y_1}
\int_0^{y_1} e^{2A}\left(y_1-y\right)^2
\,,
\label{STYW}
\end{eqnarray}
where $\Omega(y)=U(y)/U(y_1)$ and
%
$U(y)=\int_0^y h^2(y')e^{-2A(y')}$.
%
Comparing the prediction of these parameters for a fixed value of $m_{KK}$ with experimental data imposes lower bounds on the value of $m_{KK}$. We can see from Eq.~(\ref{STYW}) that the $T$ parameter is volume enhanced, the $S$ parameter is volume independent to leading order in the volume expansion while $W$ and $Y$ are volume suppressed. Therefore, we will pay attention in the following to the $T$ and $S$ parameters, and  check in all cases a posteriori that for the obtained bounds on $m_{KK}$ the values of $W$ and $Y$ are below the experimental data and hence they do not impose any further constraint~\cite{foot:3}.

The current experimental bounds on oblique observables for a SM reference Higgs mass, $m_{H,r}=117$ GeV, and assuming $U=0$ are given by 
$T = 0.07 \pm 0.08$ and
 $S = 0.03 \pm 0.09 $,
with a correlation between $S$ and $T$ of $87\%$ in the fit~\cite{Nakamura:2010zzi}.
Moreover, the one-loop contribution to the $S$ and $T$ parameters of a SM Higgs boson with a mass $m_H$, normalized to its values at the reference Higgs mass $m_{H,r}$ is given by
\begin{eqnarray}
\Delta S&=&\frac{1}{2\pi} \left[ g_S(m_H^2/m_Z^2)-g_S(m_{H,r}^2/m_Z^2) \right] \,,\nonumber\\
g_S(u)&=&\int_0^1 dx\, x(5x-3)\log(1-x+u x) \,, 
\end{eqnarray}
and~\cite{Veltman:1976rt}
\begin{eqnarray}
\Delta T&=&\frac{-3}{16\pi s^2_W}\left[ g_T(m_H^2/m_Z^2)-g_T(m_{H,r}^2/m_Z^2) \right] \,,\nonumber\\
g_T(u)&=&y\frac{\log c^2_W-\log u}{c^2_W-u}+\frac{\log u}{c_W^2(1-u)}\,.
\label{DeltaT} 
\end{eqnarray}
In the limit where the Higgs masses are much larger than $m_Z$ one recovers the approximate behavior in Ref.~\cite{Peskin:1991sw}.

We will first concentrate in the RS model with a pure AdS metric $A(y)=ky$ where the bulk Higgs mass is a constant given by $M^2=a(a-4)k^2$, and where the parameter $a$ has the simple holographic interpretation $\dim(\mathcal O_H)=a$. In this case the solution to the equations of motion is given by 
\begin{equation}
h(y)=c_1e^{aky}\left(1+c_2e^{2(2-a)ky}\right)\ ,
\label{h12}
\end{equation}
where $c_{1,2}$ depend on the model parameters. We can get the solution $h(y)\sim e^{aky}$ without any fine-tuning for $a \gtrsim 2$  since near the IR boundary where EWSB occurs $e^{2(2-a)ky_1}\ll 1$ and the second term is always irrelevant. On the contrary for $a\lesssim 2$ in Eq.~(\ref{h12}) the second term would be dominating and we get the same solution as in the previous case, i.e.~$h(y)\sim e^{a'ky}$ where $a'=4-a>2$, unless we fine-tune $c_2=0$ in which case we do not solve the hierarchy problem. Using the previous holographic interpretation of the parameter $a$ we can conclude that the hierarchy problem is solved in the dual theory provided $\dim(\mathcal O_H)>2$ as expected.

\begin{figure}[b]
\begin{psfrags}
	\input{ellipses-psfrag.tex}
	\includegraphics[width=0.45\textwidth]{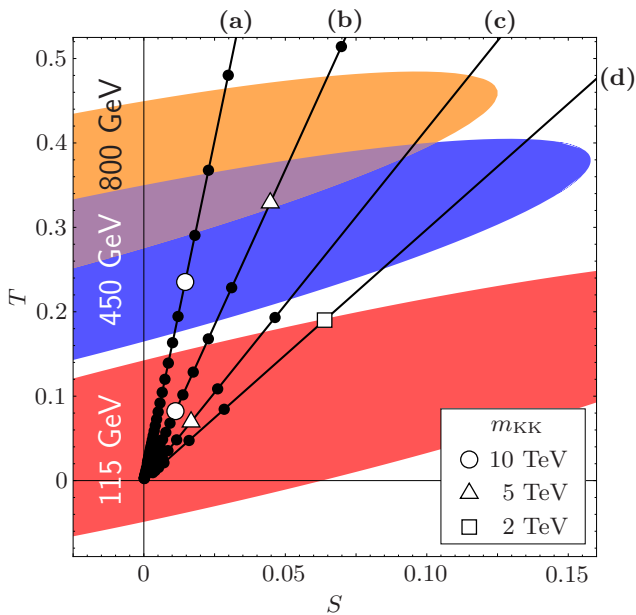}
\end{psfrags}
\caption{\label{fig:ellipses} \it 95\% CL regions in the $(S,T)$ plane for different values of the Higgs mass. Ray (a) [(b)] is RS with a localized [bulk with a=2.1] Higgs boson. Ray (c) [(d)] is model (\ref{metrica}) with $k\Delta=1$ and $\nu=0.7$ [$\nu=0.6$]. Dot spacing is 1 TeV. Increasing values of $m_{KK}$ correspond to incoming fluxes.}
\end{figure}

Using now the Higgs profile Eq.~(\ref{h12}) one obtains that $Z=1/2(a-1)<1/2$ is a small number and that the tree-level approximation on the Higgs mass is given by $m_H^2\sim\rho^2$, which points towards heavy Higgs masses unless some dimensionless constant in the brane potential is fine-tuned. The IR scale $\rho$, or the related KK-mode gauge boson mass $m_{KK}\simeq 2.4 \rho$, is bounded by EWPO. 

\begin{figure}[t]
\begin{psfrags}
	\input{boundsa-psfrag.tex}
	\includegraphics[width=0.45\textwidth]{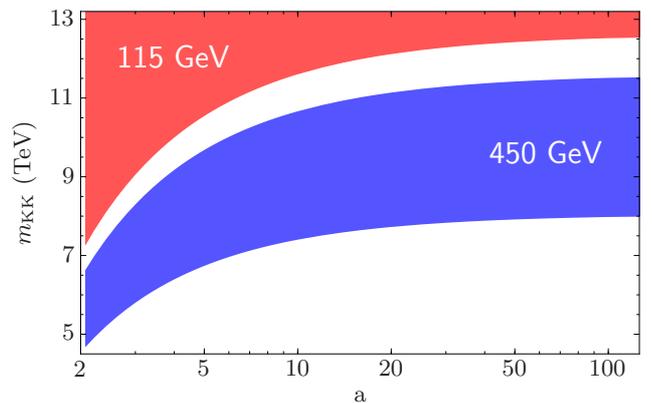}
\end{psfrags}
\caption{\label{fig:boundsa} \it 95\% CL regions in the $(a,m_{KK})$ plane for RS and different values of the Higgs mass.}
\end{figure}
\begin{figure}[b]
\begin{psfrags}
	\input{boundsRS-psfrag.tex}
	\includegraphics[width=0.45\textwidth]{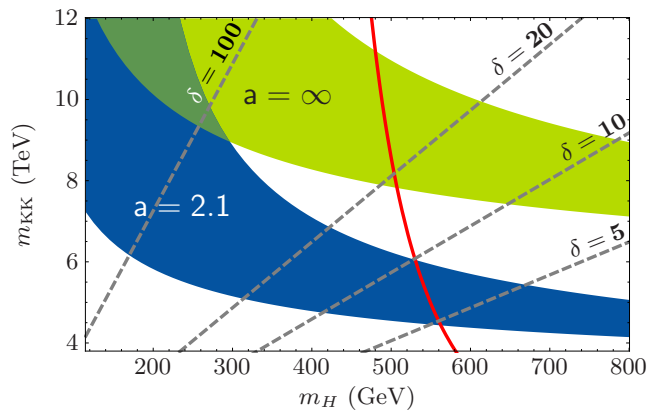}
\end{psfrags}
\caption{\label{fig:boundsRS} \it 95\% CL regions in the $(m_H,m_{KK})$ plane for RS and the cases of a localized and a bulk Higgs field with $a=2.1$. Dashed lines correspond to sensitivity $\delta=100$ (1\% fine-tuning),  $\delta=20$ (5\%), $\delta=10$ (10\%) and $\delta=5$ (20\%), for $\Lambda\sim m_{KK}$. Solid line is the perturbativity bound.}
\end{figure}

In Fig.~\ref{fig:ellipses} we show the 95\% CL ellipses in the $(S,T)$ plane for different values of the Higgs mass $m_H=115,\, 450,\, 800$ GeV. The dots correspond to different values of $m_{KK}$ and the dot spacing is 1 TeV, while the special symbols correspond to fixed values of $m_{KK}$ as indicated in the plot labels. 
The values of $m_{KK}$ increase (decrease) as the dots get closer to (further from) the origin. In this way the lower (and possibly upper) bound on $m_{KK}$ can be read from the plot for the considered values of $m_H$. Ray (a) corresponds to RS with an IR localized Higgs field. This was the case considered in Ref.~\cite{Casagrande:2008hr}. Ray (b) corresponds to RS with a bulk Higgs field where $a=2.1$. We can see from the plot that the 95\% CL window for a localized (bulk) Higgs field with e.g.~ $m_H=450$ GeV is 8 TeV $\lesssim m_{KK}\lesssim 11.6$ TeV (4.6 TeV $\lesssim m_{KK}\lesssim 6.6$ TeV). This behavior is exhibited in Fig.~\ref{fig:boundsa}, which shows the 95\% CL allowed regions in the $(a,m_{KK})$ plane and different values of the Higgs mass.

In Fig.~\ref{fig:boundsRS} the 95\% CL regions in the $(m_H,m_{KK})$ plane are exhibited for the cases of a localized and a bulk Higgs field. The solid line is the perturbativity bound defined by the condition that two-loop corrections to the $\beta_\lambda$-function be 50\% of the one-loop correction, $\beta_\lambda^{(2)}=0.5\beta_\lambda^{(1)}$~\cite{Hambye:1996wb}. The region on the right of the solid line is excluded. For a localized Higgs boson the 95\% CL lower bound on the mass of gauge KK modes is $m_{KK}> 7.8$~TeV (which corresponds to $m_H< 510$~GeV), while for a bulk Higgs boson $m_{KK}> 4.4$~TeV ($m_H< 560$~GeV).

\begin{figure}[ht]
\begin{psfrags}
	\input{alpha-psfrag.tex}
	\includegraphics[width= 0.45\textwidth]{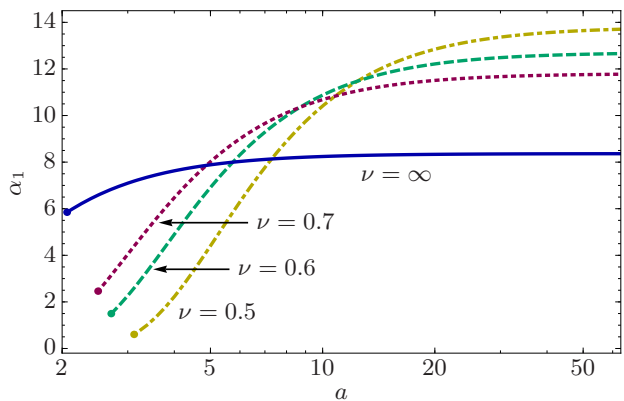}
\end{psfrags}
\caption{
\label{fig:coupling}  
\it Plot of the coupling $\alpha_1$ as a function of $a$ for $\nu=0.5,0.6,0.7$ and $\infty$ (RS).  Lines end at $a=a_0$.}
\end{figure}

Next we will consider the model analyzed in Ref.~\cite{Cabrer:2010si} with a deformation of RS in the IR region. It contains a stabilizing field $\phi$ which leads to the metric%
\begin{equation}
A(y)=ky-\frac{1}{\nu^2}\log\left(1-y/y_s\right) \,,
\label{metrica}
\end{equation}
where $\nu$ is a real parameter. The metric has a \allowbreak spurious singularity located at $y_s=y_1+\Delta$, outside the physical interval~\cite{foot:4}. 
In order to solve the hierarchy problem we fix $A_1\nobreak=\nobreak A(y_1)\sim\nobreak35$, which determines implicitely $ky_1<A_1$ in terms of the other parameters.
A suitable ($\phi$~dependent) bulk Higgs mass leads to 
\begin{equation}
h(y)=c_1 e^{a ky}\left(1+c_2 \int^ye^{4A(y')-2aky'}\right)\,,
\end{equation}
 and $h(y)\sim e^{aky}$ imposes the constraint $a\gtrsim a_0=2 A_1/ky_1$ as we analyzed in Ref.~\cite{Cabrer:2010si,Cabrer:2011mw}. There it was shown that in many cases $Z\gg 1$, which lowers the couplings $\alpha_n$ in Eq.~(\ref{couplingKK}) and correspondingly softens the bounds on $m_{KK}$ from EWPT. This behavior is exhibited in Fig.~\ref{fig:coupling}, where we plot the first mode coupling $\alpha_1$ versus $a$ for different values of $\nu$. We see from the plot that the global effect is a combination of small $a$ (less localized Higgs field) and small $\nu$ (departure from AdS in the IR). It also turns out that the main effect comes from the $1/Z$ factor in Eq.~(\ref{couplingKK}).

\begin{figure}[ht]
\begin{psfrags}
	\input{boundsMOD-psfrag.tex}
	\includegraphics[width=0.45\textwidth]{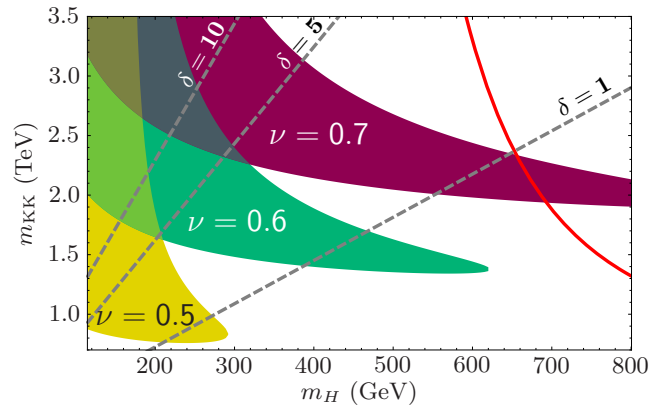}
\end{psfrags}
\caption{
\label{fig:boundsMOD}  
\it 95\% CL regions in the $(m_H,m_{KK})$ plane for 
model (\ref{metrica}) with $k\Delta=1$ and $\nu=0.7,\,0.6,\, 0.5$.  Dashed lines correspond to sensitivity $\delta=1$ (no fine-tuning), $\delta=5$ (20\%) and $\delta=10$ (10\%). Solid line corresponds to the perturbativity bound.}
\end{figure}

 In Fig.~\ref{fig:ellipses} we have considered two particularly interesting cases corresponding to $k\Delta=1$ and $\nu=0.7$ [ray (c)] and $\nu=0.6$ [ray (d)].  For case (c) we can see that for $m_H=450$ GeV the 95\% CL window for $m_{KK}$ is $2.1\, {\rm TeV}\lesssim m_{KK}\lesssim 2.9\, {\rm TeV}$ and for case (d) it is $1.4\, {\rm TeV}\lesssim m_{KK}\lesssim 1.7\, {\rm TeV}$, which are in principle accessible at the LHC energies. This shows that for model (\ref{metrica}): i) A heavy Higgs field can be consistent with KK-modes accessible at LHC energies; ii) The measurement of the Higgs mass at LHC should constrain the model parameters. These two features are exhibited in Fig.~\ref{fig:boundsMOD}, where we show the 95\% CL allowed regions in the $(m_H,m_{KK})$ plane for various values of the parameters. The solid line is the perturbativity bound and the region on its right is excluded. Then using EWPT one can extract the absolute bound on the Higgs mass as $m_H\lesssim 750$ GeV.

In conclusion, we have considered models where the 5D SM gauge bosons and the Higgs boson propagate in a warped extra dimension. In particular, we have considered the RS model and a more general 5D metric where there is an IR deformation of the conformal symmetry. We have shown that a heavy Higgs boson is more natural than a light one and moreover the EWPT can be fulfilled for lighter values of the KK masses. For the RS model and a bulk Higgs field with $m_H=450$ GeV there is a window $4.6\, {\rm TeV}\lesssim m_{KK}\lesssim 6.6\, {\rm TeV}$, while for the general model with IR deformations the KK spectrum can be accessible at the LHC. Moreover the Higgs boson discovery at the LHC will constrain the latter model parameters.

This work is supported in part by the Spanish Consolider-Ingenio 2010
Programme CPAN (CSD2007-00042) and by CICYT-FEDER-FPA2008-01430.  The work of JAC is supported by the Spanish Ministry of Education through a FPU grant. The research of GG is supported by the ERC Advanced Grant 226371, the ITN programme PITN-GA-2009-237920 and the IFCPAR CEFIPRA programme 4104-2.

\end{document}